\documentclass{aa}

\usepackage{txfonts}
\usepackage{graphicx}
\usepackage{natbib}
\bibpunct{(}{)}{;}{a}{}{,}

\begin{document}

\title{The ALHAMBRA survey\thanks{Based on observations collected at the German-Spanish Astronomical Center, Calar Alto, jointly operated by the Max-Planck-Institut f\"ur Astronomie (MPIA) at Heidelberg and the Instituto de Astrof\'{\i}sica de Andaluc\'{\i}a (CSIC)}: Estimation of the clustering signal encoded in the cosmic variance}

\author{C.~L\'opez-Sanjuan\inst{\ref{a1}} \and
A.~J.~Cenarro\inst{\ref{a1}} \and
C.~Hern\'andez-Monteagudo\inst{\ref{a1}} \and
P.~Arnalte-Mur\inst{\ref{a3}} \and
J.~Varela\inst{\ref{a1}} \and
K.~Viironen\inst{\ref{a1}} \and
A.~Fern\'andez-Soto\inst{\ref{a12},\ref{a5}} \and  
V.~J.~Mart\'inez\inst{\ref{a3},\ref{a4},\ref{a5}} \and
E.~Alfaro\inst{\ref{a2}} \and
B.~Ascaso\inst{\ref{b3}} \and  
A.~del~Olmo\inst{\ref{a2}} \and
L.~A.~D{\'{\i}}az-Garc{\'{\i}}a\inst{\ref{a1}} \and
Ll.~Hurtado-Gil\inst{\ref{a3},\ref{a12}} \and
M.~Moles\inst{\ref{a1},\ref{a2}} \and
A.~Molino\inst{\ref{b2},\ref{a2}} \and
J.~Perea\inst{\ref{a2}} \and
M.~Povi\'c\inst{\ref{a2}} \and
J.~A.~L.~Aguerri\inst{\ref{a10},\ref{a11}} \and
T.~Aparicio-Villegas\inst{\ref{b1},\ref{a2}} \and
N.~Ben\'itez\inst{\ref{a2}} \and
T.~Broadhurst\inst{\ref{a6},\ref{a7}} \and
J.~Cabrera-Ca\~no\inst{\ref{a8}} \and
F.~J.~Castander\inst{\ref{a9}} \and
J.~Cepa\inst{\ref{a10},\ref{a11}} \and
M.~Cervi\~no\inst{\ref{a2},\ref{a10},\ref{a11}} \and
D.~Crist\'obal-Hornillos\inst{\ref{a1}} \and
R.~M.~Gonz\'alez~Delgado\inst{\ref{a2}} \and
C.~Husillos\inst{\ref{a2}} \and
L.~Infante\inst{\ref{a13}} \and  
I.~M\'arquez\inst{\ref{a2}} \and
J.~Masegosa\inst{\ref{a2}} \and
F.~Prada\inst{\ref{a2},\ref{a14},\ref{a15}} \and 
J.~M.~Quintana\inst{\ref{a2}}
}

\institute{Centro de Estudios de F\'isica del Cosmos de Arag\'on, 
            Plaza San Juan 1, 44001 Teruel, Spain\\\email{clsj@cefca.es}\label{a1} 
            \and
Observatori Astron\`omic, Universitat de Val\`encia, 
            C/ Catedr\`atic Jos\'e Beltr\'an 2, E-46980, Paterna, Spain\label{a3}
            \and
Instituto de F\'isica de Cantabria (CSIC-UC), E-39005 Santander, Spain\label{a12}
            \and
Unidad Asociada Observatorio Astron\'omico (IFCA-UV), 
            E-46980, Paterna, Spain\label{a5}
            \and
Departament d'Astronomia i Astrof\'isica, 
            Universitat de Val\`encia, E-46100, Burjassot, Spain\label{a4}
            \and
IAA-CSIC, Glorieta de la Astronom\'ia s/n, 18008 Granada, Spain\label{a2} 
            \and
GEPI, Observatoire de Paris, CNRS, Universit\'e Paris Diderot, 61, 
               Avenue de l’Observatoire 75014, Paris France\label{b3}
            \and
Instituto de Astronom{\'{\i}}a, Geof{\'{\i}}sica e Ci\'encias Atmosf\'ericas, 
              Universidade de S{\~{a}}o Paulo, S{\~{a}}o Paulo, Brazil\label{b2}
            \and
Instituto de Astrof\'isica de Canarias, V\'ia L\'actea s/n, 38200 
              La Laguna, Tenerife, Spain\label{a10}
            \and
Departamento de Astrof\'isica, Facultad de F\'isica, 
              Universidad de La Laguna, 38206 La Laguna, Spain\label{a11}
            \and
Observat\'orio Nacional-MCT, Rua Jos\'e Cristino, 77. CEP 20921-400, 
              Rio de Janeiro-RJ, Brazil\label{b1}
            \and
Department of Theoretical Physics, 
            University of the Basque Country UPV/EHU, 48080 Bilbao, Spain\label{a6}
            \and
IKERBASQUE, Basque Foundation for Science, Bilbao, Spain\label{a7}
            \and
Departamento de F\'isica At\'omica, Molecular y Nuclear, 
             Facultad de F\'isica, Universidad de Sevilla, 41012 Sevilla, Spain\label{a8}
            \and
Institut de Ci\`encies de l'Espai (IEEC-CSIC), Facultat de Ci\`encies, 
             Campus UAB, 08193 Bellaterra, Spain\label{a9}
            \and
Departamento de Astronom\'ia, Pontificia Universidad Cat\'olica. 
              782-0436 Santiago, Chile\label{a13}
            \and
Instituto de F\'{\i}sica Te\'orica, (UAM/CSIC), Universidad Aut\'onoma 
              de Madrid, Cantoblanco, E-28049 Madrid, Spain \label{a14}
            \and
Campus of International Excellence UAM+CSIC, Cantoblanco, 
               E-28049 Madrid, Spain \label{a15}
}

\date{Submitted 15/06/2015 -- Accepted?}

\abstract
{}
{The relative cosmic variance ($\sigma_v$) is a fundamental source of uncertainty in pencil--beam surveys and, as a particular case of count--in--cell statistics, can be used to estimate the bias between galaxies and their underlying dark--matter distribution. Our goal is to test the significance of the clustering information encoded in the $\sigma_v$ measured in the ALHAMBRA survey.}
{We measure the cosmic variance of several galaxy populations selected with $B-$band luminosity at $0.35 \leq z < 1.05$ as the intrinsic dispersion in the number density distribution derived from the 48 ALHAMBRA subfields. We compare the observational $\sigma_v$ with the cosmic variance of the dark matter expected from the theory, $\sigma_{v,{\rm dm}}$. This provides an estimation of the galaxy bias $b$.}
{The galaxy bias from the cosmic variance is in excellent agreement with the bias estimated by two--point correlation function analysis in ALHAMBRA. This holds for different redshift bins, for red and blue subsamples, and for several $B-$band luminosity selections. We find that $b$ increases with the $B-$band luminosity and the redshift, as expected from previous work. Moreover, red galaxies have a larger bias than blue galaxies, with a relative bias of $b_{\rm rel} = 1.4 \pm 0.2$.}
{Our results demonstrate that the cosmic variance measured in ALHAMBRA is due to the clustering of galaxies and can be used to characterise the $\sigma_v$ affecting pencil--beam surveys. In addition, it can also be used to estimate the galaxy bias $b$ from a method independent of correlation functions.}


\titlerunning{The ALHAMBRA survey. Estimation of the clustering signal encoded in the cosmic variance}

\authorrunning{L\'opez-Sanjuan et al.}

\maketitle

\section{Introduction}\label{intro}
One fundamental uncertainty in any observational measurement derived from galaxy surveys is the relative cosmic variance ($\sigma_v$, also called sample variance), arising from the underlying large--scale density fluctuations and leading to variances larger than those expected from simple Poisson statistics. The most efficient way to tackle the cosmic variance is to split the survey into several independent areas on the sky. This minimises the sampling problem and it is better than increasing the volume in a wide contiguous field \citep[e.g.][]{driver10}. However, the uncertainties in many existing surveys are dominated by the cosmic variance because of the observational constraints (depth vs area). Thus, a proper estimation of $\sigma_v$ is needed to fully describe the error budget in deep cosmological surveys.

The impact of the cosmic variance in a given survey and redshift range can be estimated using two basic methods: theoretically, by analysing cosmological simulations \citep[e.g.][]{somerville04,trenti08,stringer09,moster11}, or empirically, by sampling a larger survey \citep[e.g.][]{driver10, clsj14ffcosvar, keenan14}. Unfortunately, the value of previous empirical work is limited by the precise understanding of the measured cosmic variance: the estimated $\sigma_v$ must encode, by definition, the clustering of the studied populations, but no previous work in the literature has tested this requirement.

The galaxy bias $b$ is the relationship between the spatial distribution of galaxies and the underlying dark--matter density field \citep{kaiser84,bardeen86,mo96}. The cosmic variance $\sigma_v$ is a particular case of count--in--cell statistics \citep{peebles80,efstathiou90,efstathiou95,andreani94,adelberger98,dekel99,marinoni05,robertson10,yang11_cic,diporto14}, and can be therefore used to estimate $b$ by comparison with $\sigma_{v,{\rm dm}}$, the cosmic variance of the dark matter expected from the theory. The galaxy bias computed from the cosmic variance, noted $b_v$, encodes the clustering information of $\sigma_v$ and has to be the same that the bias estimated with an independent method at similar scales. We use the bias computed from the two--point correlation function, noted $b_{\xi}$, as a benchmark because it is a well-tested method widely used in the literature. The agreement between $b_v$ and $b_{\xi}$ implies that the origin of the observational $\sigma_v$ is the clustering of galaxies, as desired, while the discrepancy reflects deficiencies in our methodology to measure $\sigma_v$.

We take advantage of the unique design, depth, and photometric redshift accuracy of the Advanced, Large, Homogeneous Area, Medium--Band Redshift Astronomical (ALHAMBRA) survey\footnote{{\tt http://www.alhambrasurvey.com}} \citep{alhambra} to study the clustering signal encoded in the cosmic variance. We compare the observed cosmic variance with the expectations from non--linear theory to estimate the galaxy bias $b_v$. Then, we compare $b_v$ with the $b_{\xi}$ reported by \citet[][AM14 hereafter]{arnaltemur14} from two--point correlation function analysis using the same data set, cosmological parameters, and theoretical expectations.

The paper is organised as follows. In Sect.~\ref{data}, we present the ALHAMBRA survey and its photometric redshifts and probability distribution functions, and in Sect.~\ref{bmeasure} our estimation of the cosmic variance and the galaxy bias. We compare our results with the bias estimated from correlation function analysis in Sect.~\ref{am14}. Finally, we present our summary and conclusions in Sect.~\ref{conclusion}. Throughout this paper we use a standard cosmology with $\Omega_{\rm m} = 0.27$, $\Omega_{\Lambda} = 0. 73$, $H_0 = 100 h$ km s$^{-1}$ Mpc$^{-1}$, and $\sigma_{8} = 0.816$. To avoid systematics in our results, these cosmological parameters are the same as in AM14. Magnitudes are given in the AB system \citep{oke83} and the absolute $B-$band magnitudes are given as $M_B - 5 \log_{10} h$, even when not explicitly indicated.

\begin{table*}
\caption{ALHAMBRA survey fields.}
\label{alhambra_fields_tab}
\begin{center}
\begin{tabular}{lcccc}
\hline\hline\noalign{\smallskip}
Field      &    Overlapping     &    RA    &    DEC     &    subfields / area \\
name       &      survey        &   (J2000) & (J2000)    &   (no. / deg$^2$)\\
\noalign{\smallskip}
\hline
\noalign{\smallskip}
ALHAMBRA-2  &  DEEP2    \citep{deep2}   & 02 28 32.0    & +00 47 00   &  8 / 0.377    \\
ALHAMBRA-3  &  SDSS     \citep{sdssdr8} & 09 16 20.0    & +46 02 20   &  8 / 0.404    \\
ALHAMBRA-4  &  COSMOS   \citep{cosmos}  & 10 00 00.0    & +02 05 11   &  4 / 0.203    \\
ALHAMBRA-5  &  GOODS-N  \citep{goods}   & 12 35 00.0    & +61 57 00   &  4 / 0.216    \\
ALHAMBRA-6  &  AEGIS    \citep{aegis}   & 14 16 38.0    & +52 24 50   &  8 / 0.400    \\
ALHAMBRA-7  &  ELAIS-N1 \citep{elais}   & 16 12 10.0    & +54 30 15   &  8 / 0.406    \\
ALHAMBRA-8  &  SDSS     \citep{sdssdr8} & 23 45 50.0    & +15 35 05   &  8 / 0.375    \\
Total       &           &               &             & 48 / 2.381\\
\noalign{\smallskip}
\hline
\end{tabular}
\end{center}
\end{table*}

\section{ALHAMBRA survey}\label{data}

The ALHAMBRA survey provides a photometric data set over 20 contiguous, equal--width ($\sim$300\AA), non--overlapping, medium--band optical filters (3500\AA -- 9700\AA) plus three standard broad--band near--infrared (NIR) filters ($J$, $H$, and $K_{\rm s}$) over eight different regions of the northern sky \citep{alhambra}. The survey has the aim of understanding the evolution of galaxies throughout cosmic time by sampling a large cosmological fraction of the universe, for which reliable spectral energy distributions (SEDs) and precise photometric redshifts ($z_{\rm p}$) are needed. The simulations of \citet{benitez09}, which relate the image depth and the accuracy of the photometric redshifts to the number of filters, have demonstrated that the filter set chosen for ALHAMBRA can achieve a photometric redshift precision that is three times better than a classical $4 - 5$ optical broad--band filter set. The final survey parameters and scientific goals, as well as the technical requirements of the filter set, were described by \citet{alhambra}. The survey has collected its data for the 20+3 optical--NIR filters in the 3.5m telescope at the Calar Alto observatory, using the wide--field camera LAICA (Large Area Imager for Calar Alto) in the optical and the OMEGA–2000 camera in the NIR. The full characterisation, description, and performance of the ALHAMBRA optical photometric system were presented in \citet{aparicio10}. A summary of the optical reduction can be found in Crist\'obal-Hornillos et al. (in prep.), the NIR reduction is reported in \citet{cristobal09}.

The ALHAMBRA survey has observed eight well--separated regions of the sky. The wide--field camera LAICA has four chips, each  with a $15\arcmin \times 15\arcmin$ field of view  (0.22 arcsec pixel$^{-1}$). The separation between chips is $13\arcmin$. Thus, each LAICA pointing provides four distinct areas in the sky, one per chip. Six ALHAMBRA regions comprise two LAICA pointings. In these cases, the pointings define two separate strips in the sky. We assumed the four chips in each LAICA pointing to be independent subfields. We summarise the properties of the seven fields included in the first ALHAMBRA data release in Table~\ref{alhambra_fields_tab}. Currently, ALHAMBRA comprises 48 subfields of $\sim183.5$ arcmin$^2$, which can be assumed to be independent for cosmic variance studies as demonstrated by \citet{clsj14ffcosvar}.

\subsection{Bayesian photometric redshifts in ALHAMBRA}
The photometric redshifts used throughout were fully presented and tested in \citet{molino13}, and we summarise their principal characteristics below.

The photometric redshifts of ALHAMBRA were estimated with \texttt{BPZ2.0}, a new version of the Bayesian Photometric Redshift \citep[\texttt{BPZ},][]{benitez00} estimator. This is a SED--fitting method based on a Bayesian inference, where a maximum likelihood is weighted by a prior probability. ALHAMBRA relied on the \texttt{ColorPro} software \citep{colorpro} to perform point spread function (PSF) matched aperture--corrected photometry, which provided both total magnitudes and isophotal colours for the galaxies. In addition, a homogeneous photometric zero--point recalibration was performed using either spectroscopic redshifts (when available) or accurate photometric redshifts from emission--line galaxies \citep{molino13}. Sources were detected in a synthetic $F814W$ filter image, noted $I$ in the following, defined to resemble the HST/$F814W$ filter. The areas of the images affected by bright stars and those with lower exposure times (e.g. the edges of the images) were masked following \citet{arnaltemur14}. The total area covered by the current ALHAMBRA data after masking is 2.38 deg$^{2}$ (Table~\ref{alhambra_fields_tab}). Finally, a statistical star/galaxy separation was encoded in the variable \texttt{Stellar\_Flag} of the ALHAMBRA catalogues, and we kept ALHAMBRA sources with $\texttt{Stellar\_Flag} \leq 0.5$ as galaxies.

The photometric redshift accuracy, as estimated by comparison with $\sim 7200$ spectroscopic redshifts ($z_{\rm s}$), was encoded in the normalised median absolute deviation (NMAD) of the photometric vs spectroscopic redshift distribution \citep{ilbert06,eazy},
\begin{equation}
\sigma_{\rm NMAD} = 1.48 \times \bigg\langle \bigg( \frac{|\,\delta_z - \langle \delta_z \rangle\,|}{1 + z_{\rm s}} \bigg) \bigg\rangle,
\end{equation} 
where $\delta_z = z_{\rm p} - z_{\rm s}$ and $\langle \cdot \rangle$ is the median operator. The fraction of catastrophic outliers $\eta$ is defined as the fraction of galaxies with $|\,\delta_z\,|/(1 + z_{\rm s}) > 0.2$. In the case of ALHAMBRA, $\sigma_{\rm NMAD} = 0.011$ for $I \leq 22.5$ galaxies with a fraction of catastrophic outliers of $\eta = 2.1$\%. We refer to \citet{molino13} for a more detailed discussion of the ALHAMBRA photometric redshifts.

\subsection{Probability distribution functions in ALHAMBRA}\label{pdfs}
This section is devoted to the description of the probability distribution functions (PDFs) of the ALHAMBRA sources. When dealing with photometric redshifts, several studies show that it is better to use the full PDF than to use only the best $z_{\rm p}$ \citep[e.g.][]{soto02,cunha09,wittman09,myers09,schmidt13,carrasco14a}. The ALHAMBRA PDFs have been successfully used to study high--redshift ($z > 2$) galaxies \citep{viironen15}, to detect galaxy groups and clusters \citep{ascaso15}, to estimate the merger fraction \citep{clsj15ffpdf}, or to improve the estimation of stellar population parameters \citep{diazgarcia15}. 

The probability that a galaxy $i$ is located at redshift $z$ and has a spectral type $T$ is $\mathrm{PDF}_{i}\,(z,T)$. This probability density function is the posterior provided by \texttt{BPZ2.0}. The probability that the galaxy $i$ is located at redshift $z$ is then
\begin{equation}
\mathrm{PDF}_{i}\,(z) = \int \mathrm{PDF}_{i}\,(z,T)\,{\rm d}T.
\end{equation}
Moreover, the total probability that the galaxy $i$ is located at $z_1 \leq z < z_2$ is
\begin{equation}
P_{i}\,(z_1,z_2) = \int_{z_1}^{z_2} \mathrm{PDF}_{i}\,(z)\,{\rm d}z.
\end{equation}
The probability density function $\mathrm{PDF}_i\,(z,T)$ is normalised to one by definition; in other words, the probability of any galaxy $i$ being found in the whole parameter space is one. Formally, 
\begin{equation}
1 = \int \mathrm{PDF}_{i}\,(z)\,{\rm d}z  = \int\!\!\!\int \mathrm{PDF}_{i}\,(z,T)\,{\rm d}T\,{\rm d}z.
\end{equation}

The $B-$band absolute magnitude of a galaxy with observed magnitude $I = 20$ and spectral type $T$ that is located at redshift $z$ is denoted as $M_B^{20}\,(z,T)$, which is also provided by \texttt{BPZ2.0}. This function condenses the information about the luminosity distance, which depends on $z$, and the $k$ correction, which depends on $z$ and $T$, needed to translate observed $I-$band magnitudes to absolute $B$-band magnitudes. Thus, the $M_B$ of the galaxy $i$ depends on its $z$ and $T$, and we have to weight $M_B^{20}\,(z,T)$ with the corresponding ${\rm PDF}_i\,(z,T)$ to obtain $M_{B,i}$. Formally, we estimate $M_B$ as a function of $z$ as
\begin{equation}
M_{B,i}\,(z) = \frac{\int M_B^{20}\,(z,T) \times {\rm PDF}_i(z,T)\,{\rm d}T}{\int {\rm PDF}_i(z,T)\,{\rm d}T} + (I_i - 20),\label{mbz}
\end{equation}
where $I_i$ is the observed $I-$band magnitude of the source.

Thanks to the probability functions defined in this section, we are able to statistically use the output of current photometric redshift codes without losing information.

\subsection{Sample selection}\label{selection}
We focus our analysis on the galaxies in the ALHAMBRA first data release\footnote{{\tt http://cloud.iaa.es/alhambra/}}. This catalogue comprises $\sim500$k sources and is complete ($5\sigma$, $3\arcsec$ aperture) for $I \leq 24.5$ galaxies \citep{molino13}.

We perform our study in a given redshift range $z \in [z_{\rm min}, z_{\rm max})$ with $B-$band luminosity selected samples. 
Because the $M_B$ of a galaxy depends on its redshift as shown in Sect.~\ref{pdfs}, we manage the galaxy samples under study with the function $\mathcal{S}$, defined as
\begin{equation}
\mathcal{S}_i\,(z, Q, M_{B}^{\rm sel}) = \left\{\begin{array}{ll}
1, & \quad {\rm if}\ M_{B,i}\,(z) + Q \times z \leq M_{B}^{\rm sel}, \\
0, & \quad {\rm otherwise},
\end{array}\right.
\label{MBsel}
\end{equation}
where $M_{B,i}\,(z)$ is the $B-$band luminosity of the galaxy $i$ from Eq.~(\ref{mbz}), $M_{B}^{\rm sel}$ is the selection magnitude of the sample, and the term $Q \times z$ accounts for the evolution of the luminosity function with redshift. This last term ensures that we explore similar $L_B/L_B^{*}$ luminosities at every $z$, where $L_B^{*}$ is the typical $B-$band luminosity at a given redshift reported by \citet{ilbert06}.

Our goal is to compare the galaxy bias derived from the cosmic variance with the bias measured from the two--point correlation function analysis performed by AM14. To minimise any possible difference between the two studies, we define the same redshift ranges and $B-$band luminosity selections as AM14. The studied samples, defined with $Q = 0.6$ to mimic AM14 selection, are summarised in the first two columns of Table~\ref{bcosvar_tab}, and cover a $B-$band luminosity from $L_B/L_B^{*} = 0.2$ to 2 at $0.35 \leq z < 1.05$. The $L_B/L_B^{*}$ values in the table are those reported by AM14.

\section{Estimation of the galaxy bias from the cosmic variance}\label{bmeasure}

In this section we present the steps we follow to estimate the galaxy bias from the cosmic variance in ALHAMBRA. First, we expose the theoretical relation between the cosmic variance and $b$ (Sec.~\ref{theo}). Then, we detail how to measure the cosmic variance from ALHAMBRA data (Sec.~\ref{obs}). Finally, we estimate the cosmic variance of the dark matter expected from theory (Sec.~\ref{bcv}).

\subsection{Theoretical link between $\sigma_v$ and $b$}\label{theo}
To connect the cosmic variance with the galaxy bias, we follow \citet{somerville04} and \citet{moster11}. The mean $\langle N \rangle$ and the variance $\langle N^2 \rangle - \langle N \rangle^2$ in the distribution of galaxies are given by the first and second moments of the probability distribution $P_N\,(V)$, which describes the probability of counting $N$ objects within a volume $V$. The relative cosmic variance is defined as
\begin{equation}
\sigma_v^2 = \frac{\langle N^2 \rangle - \langle N \rangle^2}{\langle N \rangle^2} - \frac{1}{\langle N \rangle}.\label{cosvarteo}
\end{equation}
The second term represents the correction for the Poisson shot noise. The second moment of the object counts is
\begin{equation}
\langle N^2 \rangle = \langle N \rangle^2 + \langle N \rangle + \frac{\langle N \rangle^2}{V^2} \int_{V} \xi(|{\bf r}_{\rm a} - {\bf r}_{\rm b}|)\,{\rm d}V_{{\rm a}}\,{\rm d}V_{{\rm b}},
\end{equation}
where $\xi$ is the two--point correlation function of the sample under study \citep{peebles80}. Combining this with Eq.~(\ref{cosvarteo}), the relative cosmic variance can be written as
\begin{equation}
\sigma_v^2 = \frac{1}{V^2}\,\int_{V} \xi(|{\bf r}_{\rm a} - {\bf r}_{\rm b}|)\,{\rm d}V_{{\rm a}}\,{\rm d}V_{{\rm b}}.\label{cosvarteoxi}
\end{equation}
We can approximate the galaxy correlation function in Eq.~(\ref{cosvarteoxi}) using the correlation function for dark matter $\xi_{\rm dm}$ as $\xi = b^2\,\xi_{\rm dm}$, where $b$ is the galaxy linear bias and we assume that $b$ does not depend on scale. With this definition of the correlation function, we find that
\begin{equation}
\sigma^2_v = \frac{b^2}{V^2}\,\int_{V} \xi_{\rm dm}(|{\bf r}_{\rm a} - {\bf r}_{\rm b}|)\,{\rm d}V_{{\rm a}}\,{\rm d}V_{{\rm b}} = b^2 \sigma^2_{v, {\rm dm}},
\end{equation}
where $\sigma_{v, {\rm dm}}$ is the cosmic variance of the dark matter. Finally, the galaxy bias from the cosmic variance is
\begin{equation}
b_{v}  = \frac{\sigma_v}{\sigma_{v, {\rm dm}}}.\label{bcosvar}
\end{equation}
We estimate $\sigma_v$ from ALHAMBRA data in Sect.~\ref{obs}, and $\sigma_{v, {\rm dm}}$ from theory in Sect.~\ref{bcv}.

\subsection{Measuring the cosmic variance $\sigma_v$ in ALHAMBRA}\label{obs}
In this section we define the procedure used to measure the cosmic variance from ALHAMBRA data. We estimate the number density of galaxies ($n$) in the ALHAMBRA subfield $j$ for a given $B-$band luminosity selection $M_{B}^{\rm sel}$ and redshift range $z \in [z_{\rm min}, z_{\rm max})$ as
\begin{equation}
n_{j} = \frac{N_j}{V_j} = \frac{1}{V_j}\sum_i \int_{z_{\rm min}}^{z_{\rm max}} {\rm PDF}_i\,(z) \times \mathcal{S}_i\,(z, 0.6, M_{B}^{\rm sel})\,{\rm d}z,\label{neq}
\end{equation}
where $N_j$ is the number of galaxies that fulfil the selection, and $V_j$ is the cosmic volume probed by the subfield $j$ at $z_{\rm min} \leq z < z_{\rm max}$. The index $i$ spans all the galaxies in the subfield; that is, no pre--selection of the sources is performed. A similar probabilistic approach is used by \citet{viironen15} to study the number counts of high--redshift galaxies.

\begin{figure}
\centering
\resizebox{\hsize}{!}{\includegraphics{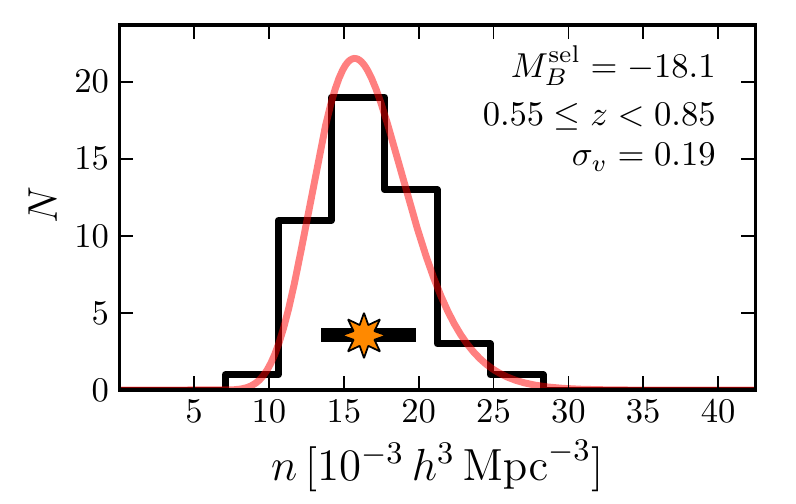}}
\resizebox{\hsize}{!}{\includegraphics{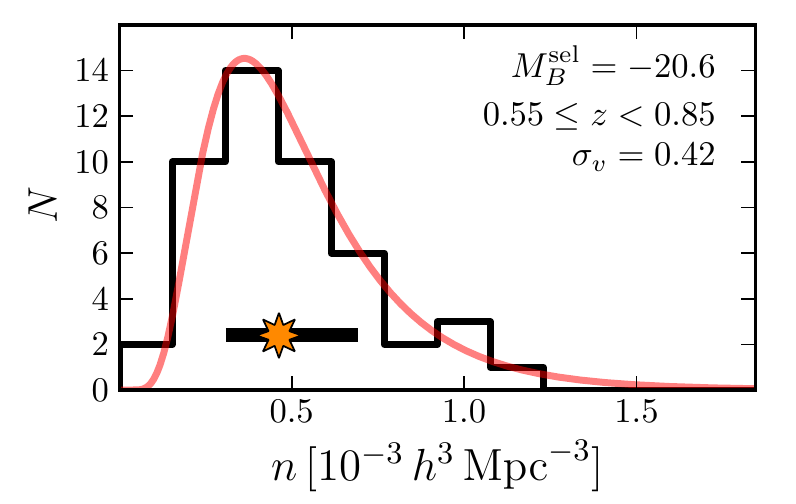}}
\caption{Number density distribution of $M_{B}^{\rm sel} = -18.1$ ({\it top panel}) and $M_{B}^{\rm sel} = -20.6$ ({\it bottom panel}) galaxies at $0.55 \leq z < 0.85$ from the 48 ALHAMBRA subfields. The star and the black bar mark the median and the intrinsic dispersion retrieved by the MLE, respectively. The red solid line shows the best MLE solution convolved with the observational errors and it is independent of the histogram binning. The derived relative cosmic variance $\sigma_v$ is labelled in the panels. A colour version of this plot is available in the electronic edition.}
\label{ng_fig}
\end{figure}

\begin{table*}
\caption{Galaxy bias estimated from the cosmic variance in ALHAMBRA.}
\label{bcosvar_tab}
\begin{center}
\begin{tabular}{cccccc}
\hline\hline\noalign{\smallskip}
      &        &    $\overline{n}$   & & & \\
$M_B^{\rm sel} -5\log_{10} h$   &   $L_{B}/L^{*}_{B}$       &  ($10^{-3}\,h^{3}$ Mpc$^{-3}$)          &    $\sigma_v$    &    $b_{v}$     &    $b_{\xi}$ \\
{\footnotesize (1)}           &   {\footnotesize (2)}       &{\footnotesize (3)}  &       {\footnotesize  (4)}          &   {\footnotesize   (5)}       &   {\footnotesize    (6)}\\
\noalign{\smallskip}
\hline
\noalign{\smallskip}
\multicolumn{6}{c}{$0.35 \leq z < 0.65$ \hspace{0.5cm} $\overline{z} = 0.52$ \hspace{0.5cm} $\sigma_{v,{\rm dm}} = 0.175$}\\
\noalign{\smallskip}
\hline
\noalign{\smallskip}
$-16.8$  &   0.16       &   $32.4 \pm 0.8$    &   $0.174 \pm 0.019$     &  $1.00 \pm 0.11$   &  $1.00 \pm 0.09$   \\
$-17.6$  &   0.27       &   $20.7 \pm 0.6$    &   $0.200 \pm 0.023$     &  $1.14 \pm 0.13$   &  $1.16 \pm 0.11$   \\
$-18.1$  &   0.37       &   $15.0 \pm 0.5$    &   $0.216 \pm 0.025$     &  $1.24 \pm 0.14$   &  $1.23 \pm 0.13$   \\
$-18.6$  &   0.50       &   $10.1 \pm 0.4$    &   $0.228 \pm 0.027$     &  $1.30 \pm 0.15$   &  $1.27 \pm 0.12$   \\
$-19.1$  &   0.73       &   $ 6.2 \pm 0.2$    &   $0.251 \pm 0.031$     &  $1.43 \pm 0.17$   &  $1.35 \pm 0.17$   \\
$-19.6$  &   1.06       &   $ 3.3 \pm 0.1$    &   $0.271 \pm 0.036$     &  $1.55 \pm 0.20$   &  $1.50 \pm 0.16$   \\
$-20.1$  &   1.52       &   $ 1.4 \pm 0.1$    &   $0.318 \pm 0.049$     &  $1.81 \pm 0.28$   &  $1.79 \pm 0.17$   \\
$-20.6$  &   2.40       &   $0.42 \pm 0.04$   &   $0.387 \pm 0.096$     &  $2.21 \pm 0.55$   &  $2.40 \pm 0.50$   \\
\noalign{\smallskip}
\hline
\noalign{\smallskip}
\multicolumn{6}{c}{$0.55 \leq z < 0.85$ \hspace{0.5cm} $\overline{z} = 0.74$ \hspace{0.5cm} $\sigma_{v,{\rm dm}} = 0.148$}\\
\noalign{\smallskip}
\hline
\noalign{\smallskip}
$-17.6$  &   0.26       &   $22.4 \pm 0.6$    &   $0.174 \pm 0.019$     &  $1.18 \pm 0.13$   &  $1.22 \pm 0.17$   \\
$-18.1$  &   0.35       &   $16.3 \pm 0.5$    &   $0.194 \pm 0.022$     &  $1.31 \pm 0.15$   &  $1.34 \pm 0.21$   \\
$-18.6$  &   0.49       &   $11.0 \pm 0.4$    &   $0.215 \pm 0.025$     &  $1.46 \pm 0.17$   &  $1.52 \pm 0.24$   \\
$-19.1$  &   0.69       &   $ 6.7 \pm 0.3$    &   $0.250 \pm 0.029$     &  $1.69 \pm 0.30$   &  $1.74 \pm 0.29$   \\
$-19.6$  &   0.97       &   $ 3.5 \pm 0.1$    &   $0.278 \pm 0.035$     &  $1.88 \pm 0.24$   &  $2.00 \pm 0.30$   \\
$-20.1$  &   1.44       &   $ 1.5 \pm 0.1$    &   $0.325 \pm 0.044$     &  $2.20 \pm 0.30$   &  $2.30 \pm 0.40$   \\
$-20.6$  &   2.13       &   $0.50 \pm 0.03$   &   $0.422 \pm 0.072$     &  $2.85 \pm 0.49$   &  $3.20 \pm 0.60$   \\
\noalign{\smallskip}
\hline
\noalign{\smallskip}
\multicolumn{6}{c}{$0.75 \leq z < 1.05$ \hspace{0.5cm} $\overline{z} = 0.91$ \hspace{0.5cm} $\sigma_{v,{\rm dm}} = 0.134$}\\
\noalign{\smallskip}
\hline
\noalign{\smallskip}
$-18.1$  &   0.34       &   $17.6 \pm 0.6$    &   $0.190 \pm 0.022$     &  $1.41 \pm 0.16$   &  $1.37 \pm 0.15$   \\
$-18.6$  &   0.46       &   $11.7 \pm 0.3$    &   $0.198 \pm 0.023$     &  $1.48 \pm 0.17$   &  $1.47 \pm 0.15$   \\
$-19.1$  &   0.66       &   $ 7.0 \pm 0.2$    &   $0.220 \pm 0.026$     &  $1.64 \pm 0.19$   &  $1.62 \pm 0.17$   \\
$-19.6$  &   0.93       &   $ 3.5 \pm 0.1$    &   $0.247 \pm 0.030$     &  $1.84 \pm 0.22$   &  $1.90 \pm 0.30$   \\
$-20.1$  &   1.35       &   $ 1.4 \pm 0.1$    &   $0.330 \pm 0.044$     &  $2.47 \pm 0.33$   &  $2.21 \pm 0.23$   \\
$-20.6$  &   1.95       &   $0.38 \pm 0.03$   &   $0.427 \pm 0.070$     &  $3.18 \pm 0.52$   &  $2.90 \pm 0.30$   \\
\noalign{\smallskip}
\hline
\end{tabular}
\end{center}
\tablefoot{Col.~(1): Absolute $B-$band magnitude at $z = 0$ used to select the sample with Eq.~(\ref{MBsel}). Col.~(2): Median luminosity of the sample in units of $L_B^{*}$ from AM14. Col.~(3): Median number density of the sample. Col.~(4): Observed cosmic variance. Col.~(5): Galaxy bias measured from the cosmic variance. Col.~(6): Galaxy bias measured from the two--point correlation function by AM14.
}
\end{table*}

The 48 number densities $n_j$ from the ALHAMBRA subfields follow a log-normal distribution,
\begin{equation}
P_{\rm LN}\,(n\,|\,\mu, \sigma) = \frac{1}{n\sqrt{2 \pi}\,\sigma}\,{\rm exp}\,\bigg[-\frac{(\ln n - \mu)^2}{2 \sigma^2}\bigg]\,,\label{Plog}
\end{equation}
where $\mu$ and $\sigma$ are the median and the dispersion of a Gaussian function in log-space,
\begin{equation}
P_{\rm G}\,(n'\,|\,\mu, \sigma) = \frac{1}{\sqrt{2 \pi}\,\sigma}\,{\rm exp}\,\bigg[-\frac{(n' - \mu)^2}{2 \sigma^2}\bigg]\,,\label{Pg}
\end{equation}
where $n' = \ln n$. This log--normal distribution of the number densities was expected because the distribution of overdense structures in the universe is described by a log--normal function \citep[e.g.][]{coles91,delatorre10,kovac10,yang11_cic}. We present two representative examples in  Fig.~\ref{ng_fig}. We checked that the number densities of all the samples under study follow a log--normal distribution with an \citet{anderson52} test. This test confirms that the distribution of the number densities is compatible with a log-normal shape and always disfavours a Gaussian distribution.

There are two origins of the observed $\sigma$: the intrinsic dispersion $\sigma_{\rm int}$ (i.e. the field-to-field variation due to the clustering of the galaxies), and the dispersion due to the Poisson shot noise $\sigma_{\rm P}$. We estimate the Poisson shot noise in each subfield as 
\begin{equation}
\sigma_{{\rm P},j} = \sqrt{N_j}/V_j
\end{equation}
and apply the maximum likelihood estimator (MLE) presented in \citet{clsj14ffcosvar} to measure the median number density of the distribution $\overline{n}$, the intrinsic dispersion $\sigma_{\rm int}$, and their uncertainties.

Applying Eq.~(\ref{cosvarteo}) to $P_{\rm LN}\,(n)$, we conclude that the observed relative cosmic variance is
\begin{equation}
\sigma^2_v = {\rm e}^{\sigma_{\rm int}^2} - 1.
\end{equation}
We estimate the $\sigma_v$ uncertainty by propagating the uncertainty in $\sigma_{\rm int}$, as estimated with the MLE \citep[see][for details]{clsj14ffcosvar}. We summarise the median number density $\overline{n}$ and the measured cosmic variance $\sigma_v$ of the studied samples in Cols. 3 and 4 of Table~\ref{bcosvar_tab}. We note that the average number densities reported by AM14 differ from our values. This discrepancy is explained by the number density definition: we estimate a probabilistic number density from Eq.~(\ref{neq}) and AM14 assumed the galaxies located at their best photometric redshift.

\subsection{Estimation of the dark--matter cosmic variance $\sigma_{v, {\rm dm}}$}\label{bcv}
The final ingredient needed to estimate the galaxy bias $b_v$ from Eq.~(\ref{bcosvar}) is $\sigma_{v, {\rm dm}}$, the cosmic variance of the dark matter expected from the theory. We calculate $\sigma_{v, {\rm dm}}$ in each volume by solving the integral in Eq.~(\ref{cosvarteoxi}) for dark matter using the code \texttt{QUICKCV}\footnote{\texttt{QUICKCV} is available at \url{www.phyast.pitt.edu/~janewman/quickcv}}, which is described in \citet{quickcv}. The code computes the cosmic variance from the dark--matter power spectrum using a window function which is 1 inside the interest volume and 0 otherwise. We obtain the dark--matter power spectrum at each redshift bin using the \texttt{CAMB} software \citep{camb}, including the non--linear corrections of \texttt{HALOFIT} \citep{halofit}.

The cosmic variance is a particular case of count--in--cell statistics, and the procedure used to estimate the galaxy linear bias is similar if either spherical or cubic volumes are used instead (see references in Sect.~\ref{intro}). In the next section, we compare the galaxy bias $b_v$ estimated from the cosmic variance with the values reported in ALHAMBRA by AM14 from two--point correlation function analysis. They estimate the galaxy bias as
\begin{equation}
b_{\xi} = \sqrt{\frac{w_{\rm p}}{w_{{\rm p},{\rm dm}}}},
\end{equation}
where $w_{\rm p}$ is the measured projected correlation function and $w_{\rm p, dm}$ is the projected dark-matter correlation function expected from the theory.


\section{Clustering signal encoded in the cosmic variance}\label{am14}
\begin{figure}[t]
\centering
\resizebox{\hsize}{!}{\includegraphics{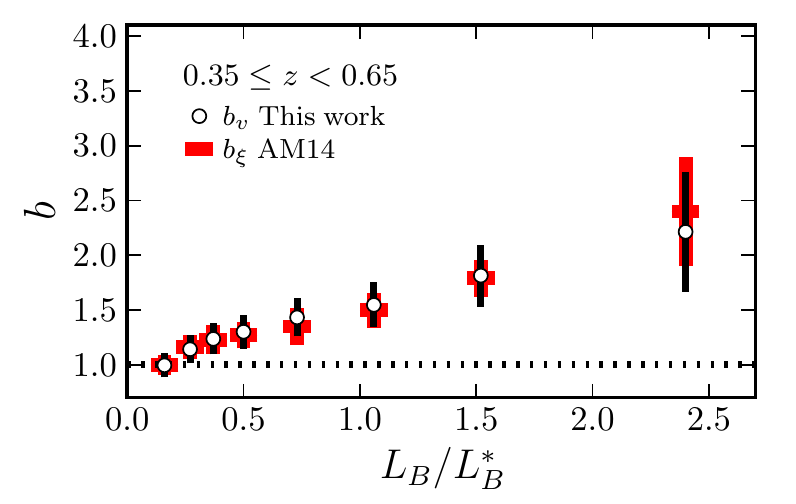}}
\resizebox{\hsize}{!}{\includegraphics{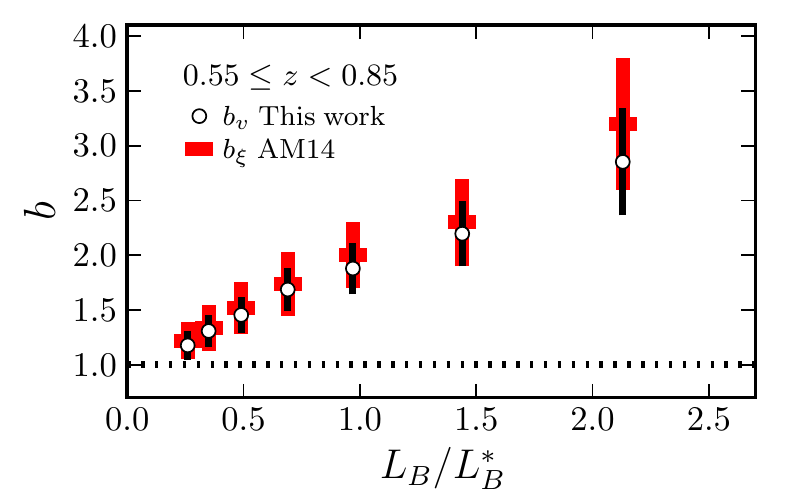}}
\resizebox{\hsize}{!}{\includegraphics{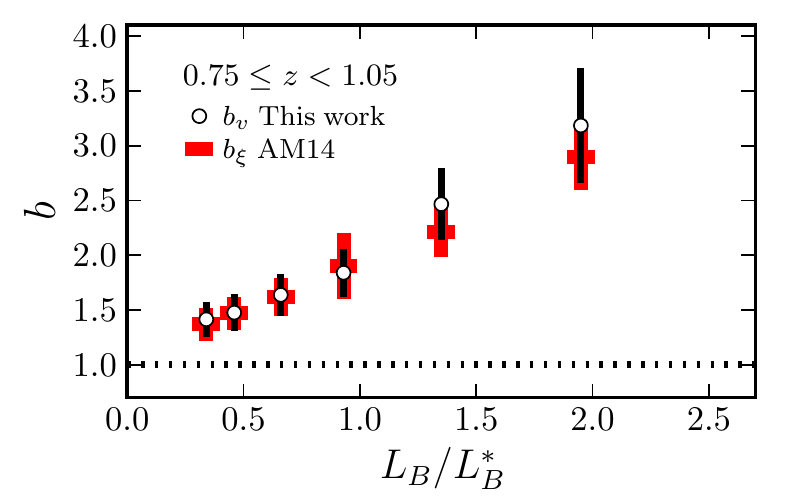}}
\caption{Galaxy bias $b$ as a function of the $B-$band luminosity in three redshift bins (labelled in the panels). The dots are from this work by cosmic variance analysis, $b_v$. The red crosses are from \citet{arnaltemur14} from the two--point correlation function analysis, $b_{\xi}$. The dotted line marks unity. A colour version of this plot is available in the electronic edition.}
\label{b_raw}
\end{figure}

We summarise the studied samples, the measured cosmic variance $\sigma_v$, the inferred galaxy bias from the cosmic variance $b_v$, and the galaxy bias $b_{\xi}$ reported by AM14 from correlation function analysis in Table~\ref{bcosvar_tab}. We show the comparison between $b_v$ and $b_{\xi}$ in three redshift bins and for several $B-$band luminosity selections in Fig.~\ref{b_raw}.

We find that the galaxy bias from the cosmic variance nicely agrees with the values from AM14, with all the estimations consistent at the $1\sigma$ level. This result has two important implications. First, it confirms that the cosmic variance measured in ALHAMBRA is due to the clustering of galaxies and the methodology applied to measure $\sigma_v$ is therefore correct. Thus, the ALHAMBRA $\sigma_v$ can be used to characterise the cosmic variance affecting pencil--beam surveys, reinforcing the results and the parametrisation of  $\sigma_v$ for close--pair studies presented by \citet{clsj14ffcosvar}. Second, we can estimate the galaxy bias $b$ from the cosmic variance, complementing  the bias measurements from correlation function analysis with an independent method. The $\sigma_v$ is measured in large volumes defined by the observational strategy of the survey, and it is unaffected by the photometric redshift precision and peculiar motions as the counts-in-cells performed with spherical or cubic volumes \citep{papa12}.

Regarding the observed trends, we find that $b_v$ increases with the $B-$band luminosity of the sample and, for a fixed selection, increases with redshift. These trends have already been found in several studies (e.g. \citealt{benoist96,norberg01,pollo06,coil06,coupon12,skibba14}; AM14). The quantitative comparison with the galaxy bias measured by other surveys is beyond the scope of this paper. Because of the excellent agreement with the AM14 values, we refer the reader to AM14 for a detailed discussion of the galaxy bias measurements in ALHAMBRA and a comparison with previous work in the literature.

In the next sections we explore the impact on the $b_v$ vs $b_{\xi}$ comparison of the ALHAMBRA fields explored (Sect.~\ref{bfields}), the power-spectrum model used (Sect.~\ref{bmodel}), the probed scale (Sect.~\ref{bscale}), and the colour of the galaxies (Sect.~\ref{bcolor}).

\subsection{Dependence on the explored fields}\label{bfields}
\begin{figure}[t]
\centering
\resizebox{\hsize}{!}{\includegraphics{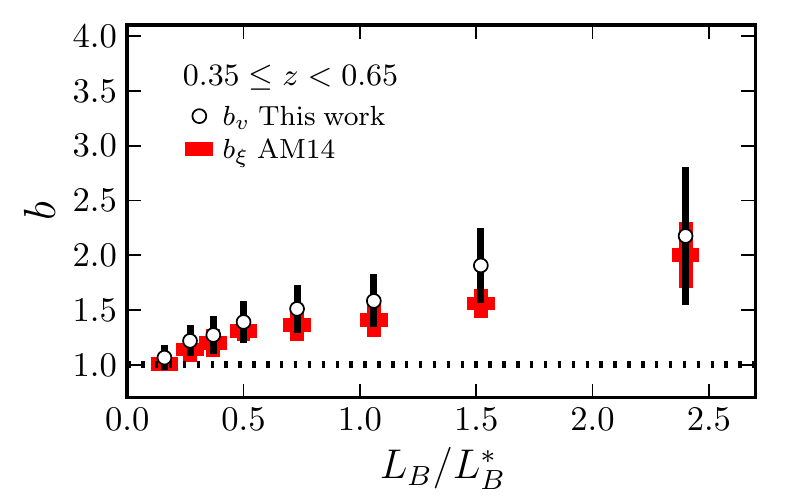}}
\resizebox{\hsize}{!}{\includegraphics{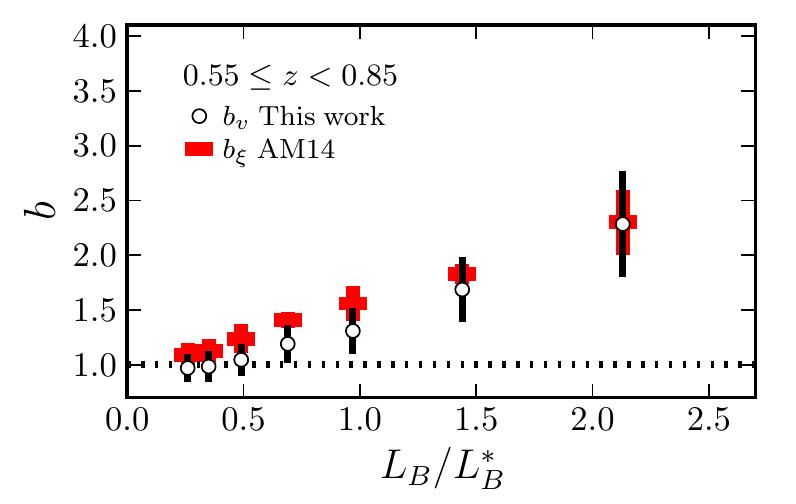}}
\resizebox{\hsize}{!}{\includegraphics{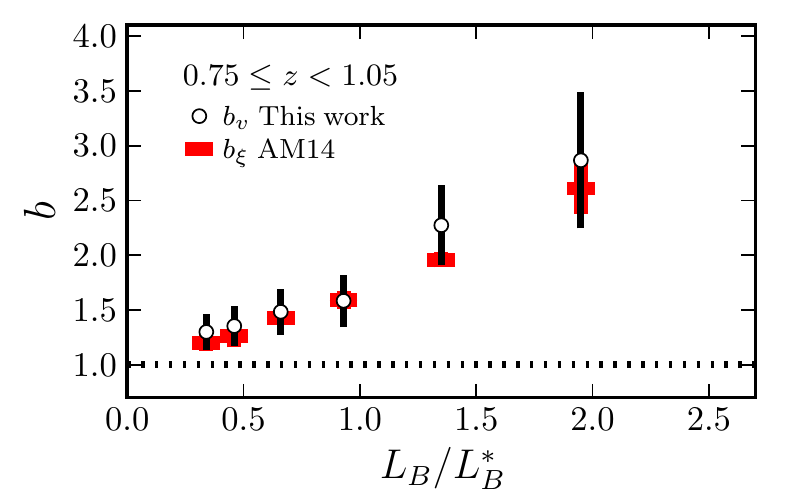}}
\caption{Galaxy bias $b$ as a function of the $B-$band luminosity in three redshift bins (labelled in the panels) with the 12 subfields from ALHAMBRA-4 and ALHAMBRA-7 removed. The dots are from this work by cosmic variance analysis, $b_v$. The red crosses are from \citet{arnaltemur14} by two--point correlation function analysis, $b_{\xi}$. The dotted line marks unity. A colour version of this plot is available in the electronic edition.}
\label{b_out}
\end{figure}

\citet{arnaltemur14} conclude that the fields ALHAMBRA-4 (COSMOS) and ALHAMBRA-7 (ELAIS--N1) have significantly different clustering properties from the other ALHAMBRA fields. Thus, in addition to the global measurements presented in the previous section, AM14 also reported the $b_{\xi}$ estimated without using the fields ALHAMBRA-4 and ALHAMBRA-7. To further check the ALHAMBRA cosmic variance, we also estimated the galaxy bias $b_v$ without the 12 subfields corresponding to the ALHAMBRA-4 and ALHAMBRA-7 fields. The results are summarised in Fig.~\ref{b_out}. As in the global case, the agreement between both estimations of the galaxy bias is remarkable and further support that the clustering measured by the correlation function is encoded in the estimated cosmic variance, as desired.

\subsection{Dependence on the power--spectrum model}\label{bmodel}
The power spectrum that we used to estimate $\sigma_{v, {\rm dm}}$ accounts for the non--linear evolution of the dark matter. The popular work of \citet{moster11} also uses \texttt{QUICKCV} but with a linear power spectrum. We checked that the values from \citet{moster11} are recovered if linear theory is used, and we estimated the impact of non--linear evolution in the theoretical models. We find that the $\sigma_{v,{\rm dm}}$ from non-linear theory is larger by 15\%, 9\%, and 6\% than the cosmic variance from linear theory reported by \citet{moster11} at $\overline{z} = 0.52$, 0.74, and 0.91, respectively. These differences are below the 20\% accuracy targeted by \citet{moster11} and decrease with redshift for two main reasons: first, the impact of the non--linear evolution increases with the growth of the structures, and  therefore decreases with $z$. Second, the non--linear effects are important at scales smaller than $\sim 1h^{-1}$ Mpc. The scale probed by each ALHAMBRA subfield increases with $z$, decreasing the impact of the non--linear evolution.

\subsection{Dependence on the probed scale}\label{bscale}
Throughout this paper we assume that the galaxy bias $b$ does not depend on scale, but $b$ is a scale--dependent parameter \citep[e.g.][]{heavens98,mann98,cresswell09}. The observational studies find that the bias is nearly constant for scales larger than $r \sim 1h^{-1}$ Mpc (the linear bias regime), with larger $b$ values at smaller scales \citep[e.g.][]{white11,parejko13}. The values of $b_{\xi}$ reported by AM14 are computed in the linear bias regime, from $1h^{-1}$ Mpc to $10h^{-1}$ Mpc. In the cosmic variance case, we integrate all the scales inside the volume of interest, and the measurements of $b_v$ are affected by scales smaller than $r = 1h^{-1}$ Mpc. The integration in Eq.~(\ref{cosvarteoxi}) weights each scale by $r^2$, so the impact of the small scales decreases with the area subtended by the ALHAMBRA subfields. This subtended area increases from $\sim3.6 \times 3.6\,h^{-2}$ Mpc$^{2}$ at $\overline{z} = 0.52$ to $\sim4.5 \times 4.5 h^{-2}$ Mpc$^{2}$ at $\overline{z} = 0.91$. We find that $b_v$ and $b_{\xi}$ are similar, implying that $b_v$ is dominated by the linear bias regime and that the ALHAMBRA subfields are large enough to reduce the scale dependence of the bias.

We further test the impact of small scales with a toy model. In our model, we assumed a scale-dependent bias of the form
\begin{equation}
b\,(r) = \left\{\begin{array}{ll}
4-2r, & \quad {\rm if}\ r < 1h^{-1}\,{\rm Mpc}, \\
2, & \quad {\rm if}\ r \geq 1h^{-1}\,{\rm Mpc},
\end{array}\right.
\end{equation}
as suggested by observations \citep{white11,parejko13}. We applied this bias model to the non--linear power spectrum and computed the cosmic variance of this biased power spectrum, noted $\sigma_v^{\rm model}$. The comparison between $\sigma_v^{\rm model}$ and $\sigma_{v, {\rm dm}}$ provides an effective bias $b_{\rm eff}$ that should be close to $b_{\rm eff} = 2$ if the signal is dominated by the linear bias regime. We find $b_{\rm eff} = 2.2, 2.1$, and $2.1$ at $\overline{z} = 0.52$, 0.74, and 0.91, respectively. This confirms that small scales should affect our measurements by less than 10\%, supporting that $b_{v}$ is comparable to the linear bias measured with $b_{\xi}$ given the geometry of the ALHAMBRA survey.

\begin{table*}
\caption{Galaxy bias of galaxies with $M_B^{\rm sel} \leq -18.6$ as a function of colour estimated from the cosmic variance in ALHAMBRA.}
\label{bcosvarcol_tab}
\begin{center}
\begin{tabular}{cccccc}
\hline\hline\noalign{\smallskip}
      &    $\overline{n}$   &   &    & \\
Redshift range   &   ($10^{-3}\,h^{3}$ Mpc$^{-3}$)       &    $\sigma_v$    &  $\sigma_{v,{\rm dm}}$ &  $b_{v}$     &    $b_{\xi}$ \\
{\footnotesize (1)}       &  {\footnotesize (2)}  &       {\footnotesize  (3)}          &   {\footnotesize   (4)}       &   {\footnotesize    (5)} &   {\footnotesize    (6)}\\
\noalign{\smallskip}
\hline
\noalign{\smallskip}
\multicolumn{6}{c}{Red templates (E/S0)}\\
\noalign{\smallskip}
\hline
\noalign{\smallskip}
$0.35 \leq z < 0.6$  &    $ 3.4 \pm 0.2$    &   $0.349 \pm 0.046$       & 0.195  & $1.84 \pm 0.26$   &  $1.69 \pm 0.08$   \\
$ 0.6 \leq z < 0.8$  &    $ 3.0 \pm 0.2$    &   $0.317 \pm 0.045$       & 0.183  & $1.78 \pm 0.26$   &  $1.55 \pm 0.13$   \\
$ 0.8 \leq z < 1.0$  &    $ 2.7 \pm 0.2$   &   $0.244 \pm 0.033$        & 0.163  & $1.52 \pm 0.21$   &  $1.83 \pm 0.11$   \\
\noalign{\smallskip}
\hline
\noalign{\smallskip}
\multicolumn{6}{c}{Blue templates (S/SB)}\\
\noalign{\smallskip}
\hline
\noalign{\smallskip}
$0.35 \leq z < 0.6$  &    $ 7.2 \pm 0.3$    &   $0.229 \pm 0.032$       &  0.195  & $1.19 \pm 0.17$   &  $1.18 \pm 0.06$  \\
$ 0.6 \leq z < 0.8$  &    $ 7.4 \pm 0.3$    &   $0.224 \pm 0.031$       &  0.183  & $1.24 \pm 0.18$   &  $1.13 \pm 0.07$  \\
$ 0.8 \leq z < 1.0$  &    $ 8.4 \pm 0.3$    &   $0.190 \pm 0.024$       &  0.163  & $1.18 \pm 0.15$   &  $1.10 \pm 0.20$  \\
\noalign{\smallskip}
\hline
\end{tabular}
\end{center}
\tablefoot{Col.~(1): Redshift range under study. Col.~(2): Median number density of the sample. Col.~(3): Observed cosmic variance. Col.~(4): Dark-matter cosmic variance from the theory. Col.~(5): Galaxy bias measured from the cosmic variance. Col.~(6): Galaxy bias measured from the two--point correlation function by \citet{lluis15}.
}
\end{table*}

\begin{figure}[t]
\centering
\resizebox{\hsize}{!}{\includegraphics{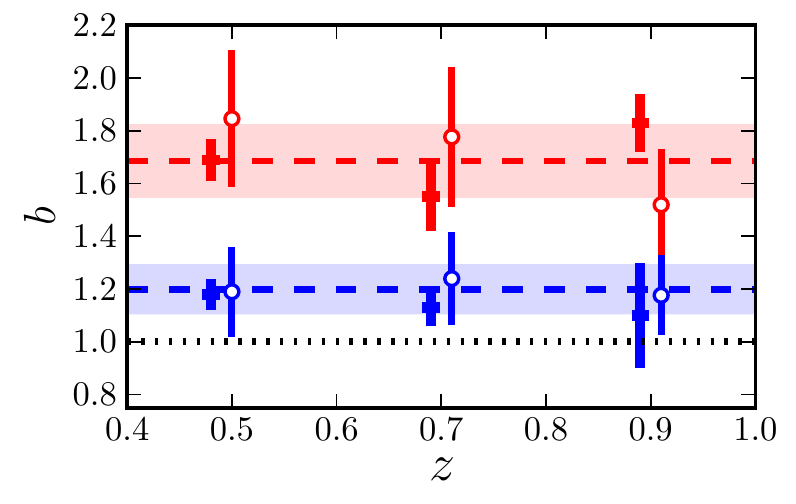}}
\caption{Galaxy bias $b$ of E/S0 (red symbols) and S/SB (blue symbols) galaxies with $M_B^{\rm sel} \leq -18.6$ as a function of redshift. The dots are from this work by cosmic variance analysis, $b_v$. The crosses are from \citet{lluis15} by two--point correlation function analysis, $b_{\xi}$. The dashed lines and the coloured areas mark the error-weighted average of $b_v$ and its error, respectively, for red ($b_{v}^{\rm red} = 1.69 \pm 0.14$) and blue ($b_{v}^{\rm blue} = 1.20 \pm 0.10$) galaxies. The dotted line marks unity. A colour version of this plot is available in the electronic edition.}
\label{b_color}
\end{figure}

\subsection{Dependence on the colour}\label{bcolor}
Following the work of AM14, \citet{lluis15} study the two-point correlation function and the galaxy bias of red and blue galaxies with $M_B^{\rm sel} \leq -18.6$ in the ALHAMBRA survey. To complete the comparison between $b_v$ and $b_{\xi}$, we also estimated the galaxy bias of red and blue galaxies with $M_B^{\rm sel} \leq -18.6$ from the cosmic variance.

To define red and blue galaxies, we take advantage of the profuse information encoded in the PDFs. Instead of selecting galaxies according to their observed colour or their best spectral template, we split each PDF into red templates ($T = {\rm E/S0}$), denoted ${\rm PDF}^{\rm red}$, and blue templates ($T = {\rm S/SB}$), denoted ${\rm PDF}^{\rm blue}$. Formally,
\begin{eqnarray}
\mathrm{PDF}\,(z) &=& \mathrm{PDF}^{\rm red}\,(z) + \mathrm{PDF}^{\rm blue}\,(z)\nonumber\\ &=& \int \mathrm{PDF}\,(z,{\rm E/S0})\,{\rm d}T + \int \mathrm{PDF}\,(z,{\rm S/SB})\,{\rm d}T.
\end{eqnarray}
In practice, the red templates have $T \in [1,5.5]$ and the blue templates have $T \in (5.5,11]$ in the ALHAMBRA catalogues. Thus, we can reliably work with colour segregations without any pre-selection of the sources \citep[e.g.][]{clsj15ffpdf}.

We summarise the basic properties of the samples under study, the measured $\sigma_v$ and $b_v$, and the $b_{\xi}$ reported by \citet{lluis15} in Table~\ref{bcosvarcol_tab}. We note that the subfields from ALHAMBRA-4 are not included in this analysis because \citet{lluis15} discard this field owing to its peculiar clustering properties. As in the general case, $b_v$ agrees with $b_{\xi}$ at the $1\sigma$ level both for red and blue galaxies, reinforcing our results and conclusions (Fig.~\ref{b_color}). We find that red galaxies are more biased than blue galaxies, with $b_v$ being compatible with a constant bias in the redshift range under study. The error-weighted averages at $0.35 \leq z < 1.0$ are $b_{v}^{\rm red} = 1.69 \pm 0.14$ and $b_{v}^{\rm blue} = 1.20 \pm 0.10$. This implies a relative bias of $b_{\rm rel} = b_{v}^{\rm red}/b_{v}^{\rm blue} = 1.4 \pm 0.2$, in agreement with previous work that find $b_{\rm rel} \sim 1.3 - 1.7$ at $z < 1$ \citep[e.g.][]{madgwick03,meneux06,coil08,delatorre11,skibba14,lluis15}.

\section{Summary and conclusions}\label{conclusion}

We estimate the significance of the clustering signal encoded in the cosmic variance $\sigma_v$ measured in the ALHAMBRA survey. With this aim, we  measured the galaxy bias $b$ from the ALHAMBRA cosmic variance. The cosmic variance is defined as the intrinsic dispersion (i.e. due to the field--to--field variations) in the number density distribution of the 48 ALHAMBRA subfields. The number densities were computed using the full probability distribution functions (PDFs) of the ALHAMBRA photometric redshifts. We compared the observed $\sigma_v$ with the cosmic variance of the dark matter expected from non--linear theory to estimate the galaxy bias $b$.

The galaxy bias from the cosmic variance nicely agrees with the bias estimated by AM14 in ALHAMBRA from two--point correlation function analysis. This confirms that the cosmic variance measured in ALHAMBRA is due to the clustering of galaxies and the methodology applied to measure $\sigma_v$ is therefore correct. Thus, the ALHAMBRA $\sigma_v$ can be used to characterise the cosmic variance affecting pencil--beam surveys, reinforcing the results and the parametrisation of the cosmic variance for close--pair studies presented by \citet{clsj14ffcosvar}. We find that the bias increases with the $B-$band luminosity of the sample and, for a fixed selection, increases with redshift, in agreement with previous studies (e.g. \citealt{norberg01,coupon12}; AM14). Moreover, red galaxies have a larger bias than blue galaxies, with a relative bias of $b_{\rm rel} = 1.4 \pm 0.2$ \citep[e.g.][]{madgwick03,meneux06,skibba14}.

The technique outlined in this paper can be used to estimate the galaxy bias $b$ of high--redshift galaxies and of different galaxy populations at $z < 1$ (Sect.~\ref{bcolor}) simply from the dispersion of the observed number densities, complementing   the bias measurements from correlation function analysis with an independent method. These number densities are measured in large volumes defined by the observational strategy of the survey rather than in the common spherical or cubic volumes, which are affected by the photometric redshift precision and peculiar motions \citep{papa12}. In addition to its cosmological and observational value, the study of the number density distribution can be useful in order to detect systematics and photometric inhomogeneities in the next generation of large photometric surveys, such as the Dark Energy Survey (DES, \citealt{des}), the Javalambre--Physics of the accelerating universe Astrophysical Survey (J--PAS, \citealt{jpas}), and the Large Synoptic Survey Telescope (LSST, \citealt{lsst}).

\begin{acknowledgements}
We dedicate this paper to the memory of our six IAC colleagues and friends who
met with a fatal accident in Piedra de los Cochinos, Tenerife, in February 2007,
with special thanks to Maurizio Panniello, whose teachings of \texttt{python}
were so important for this paper. We thank R.~Angulo, S.~Bonoli, A.~Ederoclite, A.~Orsi, and all the CEFCA staff for useful and productive discussions. We thank the anonymous referee for his/her suggestions, and K.~Xu for the pertinent comments that motivated this work.\\
This work has been mainly funded by the FITE (Fondos de Inversiones de Teruel) and the projects AYA2012-30789, AYA2006-14056, and CSD2007-00060. We also acknowledge support from the Spanish Ministry for Economy and Competitiveness and FEDER funds through grants AYA2010-15081, AYA2010-15169, AYA2010-22111-C03-01, AYA2010-22111-C03-02, AYA2011-29517-C03-01, AYA2012-39620, AYA2013-40611-P, AYA2013-42227-P, AYA2013-43188-P, AYA2013-48623-C2-1, AYA2013-48623-C2-2, ESP2013-48274, AYA2014-58861-C3-1, Arag\'on Government Research Group E103, Generalitat Valenciana projects Prometeo 2009/064 and PROMETEOII/2014/060, Junta de Andaluc\'{\i}a grants TIC114, JA2828, P10-FQM-6444, and Generalitat de Catalunya project SGR-1398.\\
A.~J.~C. and C.~H.-M. are {\it Ram\'on y Cajal} fellows of the Spanish government. A.~M. acknowledges the financial support of the Brazilian funding agency FAPESP (Post-doc fellowship - process number 2014/11806-9). M.~P. acknowledges financial support from JAE-Doc program of the Spanish National Research Council (CSIC), co-funded by the European Social Fund.\\
This research made use of \texttt{Astropy}, a community-developed core \texttt{Python} package for Astronomy \citep{astropy}, and \texttt{Matplotlib}, a 2D graphics package used for \texttt{Python} for publication-quality image generation across user interfaces and operating systems \citep{pylab}.

\end{acknowledgements}

\bibliography{biblio}
\bibliographystyle{aa}

\end{document}